\documentclass[11pt]{article}
\usepackage{amsmath,paralist,amsthm,amssymb}
\usepackage{cite}
\usepackage{algorithm}
\usepackage{algorithmic}

\usepackage[matrix,frame,arrow]{xy}
\usepackage{amsmath}

\newcommand{\ket}[1]{\left\vert{#1}\right\rangle}
\newcommand{\qw}[1][-1]{\ar @{-} [0,#1]}
\newcommand{\qwx}[1][-1]{\ar @{-} [#1,0]}

\newcommand{\gate}[1]{*{\xy *+<.6em>{#1};p\save+LU;+RU **\dir{-}\restore\save+RU;+RD **\dir{-}\restore\save+RD;+LD **\dir{-}\restore\POS+LD;+LU **\dir{-}\endxy} \qw}

\newcommand{\control}{*!<0em,.025em>-=-{\bullet}}

\newcommand{\ctrl}[1]{\control \qwx[#1] \qw}

\newcommand{\targ}{*!<0em,.019em>=<.79em,.68em>{\xy {<0em,0em>*{} \ar @{ - } +<.4em,0em> \ar @{ - } -<.4em,0em> \ar @{ - } +<0em,.36em> \ar @{ - } -<0em,.36em>},<0em,-.019em>*+<.8em>\frm{o}\endxy} \qw}

\newcommand{\gategroup}[6]{\POS"#1,#2"."#3,#2"."#1,#4"."#3,#4"!C*+<#5>\frm{#6}}

\newcommand{\lstick}[1]{*!R!<.5em,0em>=<0em>{#1}}

\newcommand{\Qcircuit}[1][0em]{\xymatrix @*[o] @*=<#1>}

\newcommand{\mc}[1]{\mathcal{#1}}
\newcommand{\mbf}[1]{\mathbf{#1}}
\newcommand{\C}{\mathbb{C}}
\newcommand{\F}{\mathbb{F}}
\newcommand{\nix}[1]{}
\newcommand{\ol}{\overline}

\newtheorem{theorem}{Theorem}
\newtheorem{corollary}[theorem]{Corollary}
\newtheorem{lemma}[theorem]{Lemma}

\newtheorem{remark}[theorem]{Remark}
\newtheorem{example}[theorem]{Example}
\newtheorem{defn}[theorem]{Definition}

\begin{document}
\title{ \Large \textbf{Encoding Subsystem Codes}}
\author{Pradeep Kiran Sarvepalli and Andreas Klappenecker\\
Department of Computer Science\\
Texas A\&M University, College Station, TX 77843\\
Email: \{pradeep,klappi\}@cs.tamu.edu
}
\date{}
\maketitle
\begin{abstract}
\noindent 
In this paper we investigate the encoding of operator quantum error correcting codes
{\em i.e.} subsystem codes. We show that encoding of subsystem
codes can be reduced to encoding of a related stabilizer code making it possible
to use all the known results on encoding of stabilizer codes. Along the way we also
show how Clifford codes can be encoded. We also show that gauge qubits 
can be exploited to reduce the encoding complexity.
\end{abstract}

\paragraph{Introduction.}
In this paper we investigate encoding of subsystem codes. 
Our main result is that encoding of a subsystem code can 
be reduced to the encoding of a related stabilizer code, thereby making use of the 
previous theory on encoding stabilizer codes \cite{cleve97,gottesman97, grassl03}. 
We shall prove this in two steps. First,
we shall show that Clifford codes can be encoded using the same methods used for stabilizer codes.
Secondly, we shall show how these methods can be adapted to encode Clifford subsystem codes.
Since subsystem codes subsume  stabilizer codes, noiseless subsystems and decoherence free
subspaces, these results imply that we can essentially use the
same methods to encode all these codes. 
In fact, while the exact details were not provided, it was suggested in \cite{poulin06} that encoding
of subsystem codes can be achieved by Clifford unitaries. Our treatment is comprehensive and gives 
 proofs for  all the claims. 

Subsystem codes can potentially lead to simpler  error recovery schemes. 
In a similar vein, they can also simplify 
the encoding process, though perhaps not as dramatically\footnote{In general, decoding  
is usually of greater complexity than encoding and for this reason it is often neglected
in comparison. This parallels the classical case where also the decoding is studied much
more extensively than encoding.}. 
These simplifications have not been investigated thoroughly, neither have 
the gains in encoding been fully characterized. Essentially, these gains are in 
two forms. In the encoded state there need not exist a one to one correspondence between
the gauge qubits and the physical qubits. However, prior to encoding such a
correspondence exists. 
We can exploit this identification between the virtual qubits and the physical
qubits before encoding to tolerate errors on the gauge qubits, a fact which
was recognized in \cite{poulin06}. Alternatively, 
we  can optimize the encoding circuits by eliminating certain encoding operations. 
The encoding operations that are saved correspond to the encoded operators on the gauge
qubits. This is a slightly subtle point and will be elaborated at length subsequently.  
We argue that optimizing the encoding circuit for the latter is much more beneficial than
simply allowing for random initialization of gauge qubits. 

{\em Notation.} 
We shall denote a finite field with $q$ elements by $\F_q$. 
Following standard convention we use $[[n,k,d]]_q$ for stabilizer codes
and $[[n,k,r,d]]_q$ for subsystem codes. The inner product of two characters of 
a group $N$, say  $\chi$ and $\theta$, is defined as $(\chi,\theta)_N=1/{|N|} \sum_{ n\in N}\chi(n)\theta(n^{-1})$. We shall denote the center of a group $N$ by $Z(N)$. 
Given a subgroup  $N\le E$, we shall denote the centralizer of $N$ in $E$ by $C_E(N)$. 
Given a matrix $A$, we consider another matrix $B$ obtained from $A$ 
by column permutation $\pi$ as being equivalent and denote this by 
$B=_{\pi} A $. Often we shall represent the basis of a group by the
rows of a matrix. In this case we will regard another basis obtained
by any row operations or permutations as being equivalent and by a slight
abuse of notation continue to denote $B=_{\pi}A$. 
The commutator of two operators  $ A$, $B$ is defined as $[A,B] = AB-BA$.

\paragraph{Encoding Stabilizer Codes \cite{cleve97,gottesman97}.} 
We shall now briefly, review the standard form encoding of stabilizer codes, 
due to Cleve and Gottesman, see \cite{cleve97,gottesman97}.
Recall the Pauli matrix operators\footnote{We consider the real version of the Pauli group in this paper.},
\begin{eqnarray}
X = \left[\begin{array}{cc}0 &1\\1&0\end{array} \right], \quad
Z = \left[\begin{array}{cc}1 &0\\0&-1\end{array} \right], \quad 
Y = \left[\begin{array}{cc}0 &-1\\1&0\end{array} \right] = XZ.
\end{eqnarray}
Let $\mc{P}_n$ be the Pauli group on $n$ qubits. An element 
element $e=  (-1)^cX^{a_1}Z^{b_1}\otimes \cdots \otimes X^{a_n}Z^{b_n}$
in $\mc{P}_n$, can be mapped to $\F_2^{2n}$ by 
 $\tau : \mc{P}_n \rightarrow \F_2^{2n}$  as
\begin{eqnarray}
\tau(e) =  (a_1,\ldots, a_n|b_1,\ldots, b_n).\label{eq:iso}
\end{eqnarray}

Given an $[[n,k,d]]_2$ code with stabilizer $S$, we can associate to $S$ (and therefore to the code), a matrix in $\F_2^{(n-k)\times 2n}$ obtained by taking the image of
any set of its generators under the mapping $\tau$. We shall refer to this 
matrix as the  \textsl{stabilizer matrix}. \index{stabilizer!matrix}
We shall refer to the stabilizer as well as any
set of generators as the stabilizer. Additionally, because of the mapping $\tau$,
we shall refer to the stabilizer matrix or any matrix obtained 
from it by row reduction or column permutations also as the stabilizer.
The stabilizer matrix can be put in the so-called ``standard form'', 
see \cite{cleve97, gottesman97}. This form also allows us to compute the encoded operators 
for the stabilizer code. 
Recall that the encoded operators allow us to 
perform computations on the encoded data without having to decode the data and then
compute.

\begin{defn}[Encoded operators]\index{encoded operators}\index{logical operators}
Given a $[[n,k,d]]_2$ stabilizer code with stabilizer $S$,
let $\ol{X}_i$, $\ol{Z}_i$ for $1\leq i\leq k$ be a set of $2k$ linearly 
independent operators in $C_{\mc{P}_n}(S)\setminus S Z(\mc{P}_n)$. The operators
$\ol{X}_i$, $\ol{Z}_i$ are said to be encoded operators for the code 
if they satisfy the following requirements. 
\begin{compactenum}[i)]
\item $[\ol{X}_i,\ol{X}_j] =0$
\item $[\ol{Z}_i,\ol{Z}_j]=0$
\item $[\ol{X}_i,\ol{Z}_j] = 2\delta_{ij}\ol{X}_i\ol{Z}_i$
\end{compactenum}
\end{defn}
The operators $\ol{X}_i$ and $\ol{Z}_j$ are referred to as encoded or logical $X$ and $Z$ operators
on the $i$th and $j$th logical qubits, respectively.
The choice of which of the $2k$ linearly independent elements of $C_{\mc{P}_n}(S)\setminus SZ(\mc{P}_n)$
we choose to call encoded $X$ operators and $Z$ operators is arbitrary;
as long as  the generators satisfy the conditions above, any choice is valid. 
Different choices lead to different sets of encoded logical states; alternatively,
a different orthonormal basis for the codespace.

\begin{lemma}[Standard form of stabilizer matrix \cite{cleve97,gottesman97}]\label{lm:stabStdForm}
\index{encoding!standard form}
Up to a permutation $\pi$, the 
stabilizer matrix of an $[[n,k,d]]_2$ code can be put in the 
following form, 
\begin{eqnarray}
S=_{\pi}\left[\begin{array}{ccc|ccc}
I_{s'}& A_1 & A_2 & B & 0 & C\\
0 & 0 & 0& D&I_{n-k-s'} & E
\end{array} \right],\label{eq:stdForm}
\end{eqnarray}
while the associated encoded operators can be derived as 
\begin{eqnarray}
\left[\begin{array}{c}
\ol{Z}\\
\ol{X} 
\end{array}\right]
=_{\pi}\left[\begin{array}{ccc|ccc}
0 & 0 &0 &A_2^t&0 &I_{k}\\
\hline
0 & E^t &I_{k}&C^t&0 &0 
\end{array}\right].\label{eq:encOps}
\end{eqnarray}
\end{lemma}
\begin{remark}Encoding 
using essentially same ideas is possible even if the  
identity matrices ($I_{s'}$ in the stabilizer matrix or $I_{k}$ in the encoded operators) 
are replaced by upper triangular matrices.
\end{remark}
The standard form of the stabilizer matrix prompts us to distinguish between
two types of the generators for the stabilizer as they affect the encoding
in different ways (although it can be shown that they are of equivalent complexity). 

\begin{defn}[Primary generators]
A generator $G_i=(a_1,\ldots,a_n|b_1,\ldots,b_n)$ with at least one nonzero 
$a_i$ is  called a primary generator.
\end{defn}
In other words, primary generators contain at least one $X$ or $Y$
operator on some qubit. 
The primary generators determine to a large extent the complexity of the 
encoding circuit along with
the encoded $X$ operators. 
The operators $\ol{X}$ are also called
seed generators and they also figure in the encoding circuit.
The encoded $Z$
operators do not.

\begin{defn}[Secondary generators]
A generator of the form $(0,\ldots,0|b_1,\ldots,b_n)$ is called
secondary generator.
\end{defn}
In the standard form encoding, the complexity of the encoded $X$ operators is determined 
by the secondary generators. 
Therefore they indirectly contribute\footnote{Indirect because the submatrix 
$E$,  figures in both 
the secondary generators, see equation~(\ref{eq:stdForm}), and also the encoded 
$X$ operators, see equation~(\ref{eq:encOps}).} to  the complexity of encoding.

We mentioned earlier that different choices of the encoded operators amounts to 
choosing different orthonormal basis for the codespace. 
However, the choice in Lemma~\ref{lm:stabStdForm} is particularly 
suitable for encoding. We can represent our input in the form 
$\ket{0}^{\otimes^{n-k}}\ket{\alpha_1\ldots\alpha_k}$ which allows us to make
the identification that $\ket{0}^{\otimes^n}$ is mapped to $\ket{\ol{0}}$, 
the logical all zero code word. 
This state is precisely the state stabilized by the stabilizer generators and 
logical $Z$ operators, (which in Lemma~\ref{lm:stabStdForm} can be seen to be consisting
of only $Z$ operators).
Given the stabilizer matrix in the standard form and the encoded operators as in 
Lemma~\ref{lm:stabStdForm}, the encoding circuit is given as follows. 
\begin{lemma}[Standard form encoding of stabilizer codes \cite{cleve97,gottesman97}]\label{lm:implement}
\index{encoding!standard form}
Let $S$ be the  stabilizer matrix of an $[[n,k,d]]_2$ stabilizer code 
in the standard form {\em i.e.}, as in equation~(\ref{eq:stdForm}).
Let $G_i$ denote the $i$th primary generator of $S$ and $\ol{X_j}$ denote the $jth$ encoded
$X$ operator as in equation~(\ref{eq:encOps}). Then these operators are in the form\footnote{We allow some
freedom in the primary generators, in that instead of $I_{s'}$ in equation~(\ref{eq:stdForm}), we
allow it be an upper triangular matrix also.}
\begin{eqnarray*}
G_i&=&(0,0,\ldots,1,a_{i+1},\ldots, a_n|b_1,\ldots,b_{s'},0,\ldots,0,b_{n-k+1},\ldots,b_n),\\
\ol{X}_j&=&(0,\ldots,0,c_{s'+1},\ldots,c_{n-k}0,\ldots,0,1=c_{n-k+j},0,\ldots,0|d_1,\ldots,d_{s'},0,\ldots,0).
\end{eqnarray*}
To encode the stabilizer code we implement the following circuits
corresponding to each of the primary generators and the encoded operators.
The generator $G_i$ is implemented after $G_{i+1}$. 
The encoded operators precede the primary generators
in their implementation but we can implement $\ol{X}_j$ before or after $\ol{X}_{j+1}$.
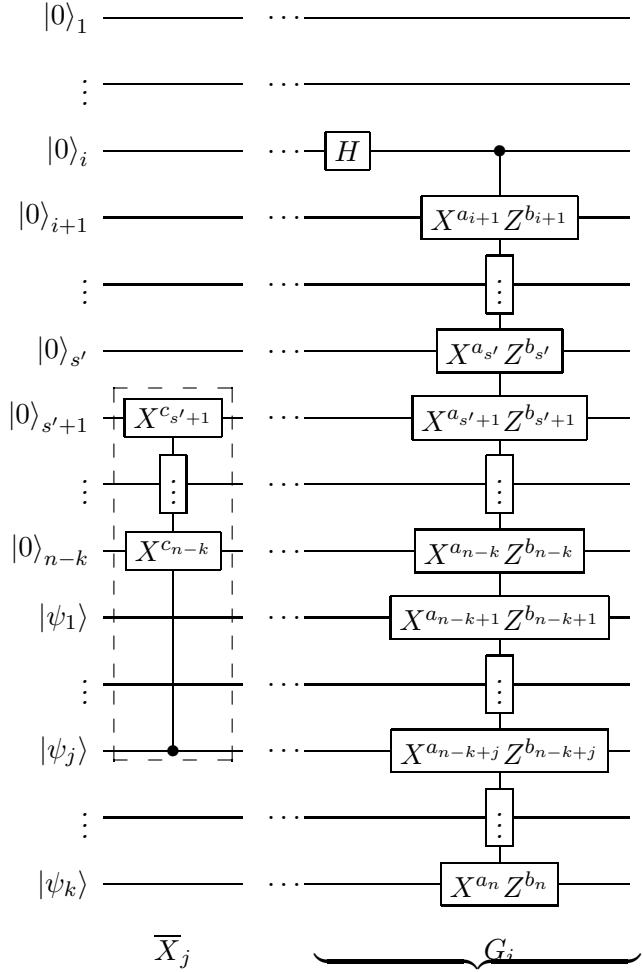
\[
\Qcircuit @C=.7em @R=.3em @!R {
\lstick{\ket{0}_1} & \qw & \qw & &\dots & & \qw& \qw & \qw &\\
\lstick{\vdots} & \qw & \qw & & \dots& & \qw&\qw & \qw& \\
\lstick{\ket{0}_i} & \qw & \qw &  &\dots& &\gate{H}& \ctrl{1} &\qw& \\
\lstick{\ket{0}_{i+1}} & \qw & \qw & &\dots & & \qw&\gate{X^{a_{i+1}}Z^{b_{i+1}}} \qwx[1]& \qw & \\
\lstick{\vdots} & \qw & \qw &  &\dots&& \qw&\gate{ \vdots }\qwx[1] & \qw \\
\lstick{\ket{0}_{s'}} & \qw & \qw & &\dots && \qw& \gate{X^{a_{s'}}Z^{b_{s'}}} \qwx[1]& \qw \\
\lstick{\ket{0}_{s'+1}} &  \gate{X^{c_{s'+1}}}\qwx[1]& \qw & &\dots& &\qw&  \gate{X^{a_{s'+1}}Z^{b_{s'+1}}} \qwx[1]& \qw \\
\lstick{\vdots} & \gate{ \vdots }  \qwx[1] & \qw &  &\dots&& \qw&\gate{ \vdots }\qwx[1] & \qw \\
\lstick{\ket{0}_{n-k}} &  \gate{X^{c_{n-k}}} & \qw & & \dots & &\qw&\gate{X^{a_{n-k}}Z^{b_{n-k}}} \qwx[1]& \qw & \\
\lstick{\ket{\psi_1}}  & \qw & \qw &  &\dots & & \qw&\gate{X^{a_{n-k+1}}Z^{b_{n-k+1}}} \qwx[1]& \qw & \\
\lstick{\vdots} & \qw & \qw & & \dots&& \qw&\gate{ \vdots } \qwx[1]& \qw \\
\lstick{\ket{\psi_j}}  & \ctrl{-3}& \qw &  &\dots && \qw&\gate{X^{a_{n-k+j}}Z^{b_{n-k+j}}}\qwx[1] & \qw \\
\lstick{\vdots} & \qw & \qw & &\dots && \qw&\gate{ \vdots } \qwx[1]& \qw \gategroup{7}{2}{12}{2}{.7em}{--}\\
\lstick{\ket{\psi_k}}  & \qw & \qw&  &\dots&& \qw&\gate{X^{a_{n}}Z^{b_{n}}}& \qw \gategroup{3}{7}{15}{9}{.7em}{_\}}& \\
 & \ol{X}_j  &  & & && &G_i&&&&\\
}
\]
\end{lemma}

To encode a stabilizer code, we first put the stabilizer matrix in the standard form,
then implement the seed generators i.e., the encoded $X$ operators, followed by the 
primary generators $i=s'$ to $i=1$ as per Lemma~\ref{lm:implement}. The complexity of
encoding the $ith$ primary generator is at most $n-i$ two qubit gates and one $H$ gate.
The complexity of encoding an encoded operator is at most $n-k-s'$ CNOT gates. This means
the complexity of standard form encoding  is upper bounded by $(2n-1-k-s')s'/2$ two 
qubit gates and $s'$ Hadamard gates; $O(n(n-k))$ gates. 
Perhaps an example will help at this juncture. 
\begin{example}
Let us consider the  $[[5,1,3]]$ code with following stabilizer. 
\begin{eqnarray*}
S & =  &\left[ \begin{array}{ccccc}
	X&I&X&X&X\\
	I&X&Z&X&Y\\
	Z&I&Z&Z&Z\\
	I&Z&Y&Z&X
\end{array}\right]
\end{eqnarray*}
The associated stabilizer matrix is given by 
\begin{eqnarray*}
S & =  &\left[ \begin{array}{ccccc|ccccc}
	1&0&1&1&1&0&0&0&0&0\\
	0&1&0&1&1&0&0&1&0&1\\
	0&0&0&0&0&1&0&1&1&1\\
	0&0&1&0&1&0&1&1&1&0 
\end{array}\right]
\end{eqnarray*}
Writing $S$ in standard form we get
\begin{eqnarray*}
S&=&\left[ \begin{array}{ccccc|ccccc}
	1&0&0&1&0&1&1&0&0&1\\
	0&1&0&1&1&0&0&1&0&1\\
	0&0&1&0&1&1&1&0&0&1\\ 
	0&0&0&0&0&1&0&1&1&1
\end{array}\right] =\left[\begin{array}{c} G_1\\G_2\\G_3\\G_4\end{array} \right].
\end{eqnarray*}
The encoded operators for this code are 
\begin{eqnarray*}
\left[ \begin{array}{c}\ol{Z}\\\ol{X}
\end{array}\right]&=&\left[ \begin{array}{ccccc|ccccc}
	0&0&0&0&0&0&1&1&0&1\\
	0&0&0&1&1&1&1&1&0&0
\end{array}\right].
\end{eqnarray*}
The stabilizer matrix has three primary generators. By Lemma~\ref{lm:implement} the encoding 
circuit is given by 
\begin{figure}[htb]
\[
\Qcircuit @C=.7em @R=.4em @! {
\lstick{\ket{0}} &\qw&\qw& \qw &\qw &\qw&\gate{H}& \ctrl{1} & \qw \\
\lstick{\ket{0}} &\qw&\qw&\qw& \gate{H} & \ctrl{1}&\qw & \gate{Z}\qwx[2] & \qw & \\
\lstick{\ket{0}} &\qw&\gate{H}& \ctrl{2} & \qw&\gate{Z}\qwx[1]&\qw & \qw& \qw & \\
\lstick{\ket{0}}&\gate{X}& \qw& \qw &\qw& \gate{X}\qwx[1] &\qw& \gate{X}\qwx[1] &\qw \\
\lstick{\ket{\psi}}&\ctrl{-1}&\qw& \gate{Y}&\qw & \gate{Y} & \qw&\gate{Z} &\qw \\
&\ol{X}  \gategroup{4}{2}{5}{2}{.7em}{--}&& G_3  \gategroup{3}{3}{5}{4}{.7em}{_\}}&& G_2  \gategroup{2}{5}{5}{6}{.7em}{_\}}& &G_1 \gategroup{1}{7}{5}{8}{.7em}{_\}}\\
}
\]
\caption{Encoding for the $[[5,1,3]]$ code}
\end{figure}
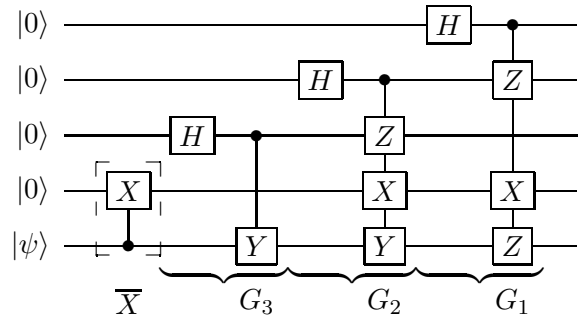
\end{example}
\begin{remark}
At this point we want to highlight that in Lemma~\ref{lm:implement}, we let
primary generators to be in the upper triangular form instead of
the standard form given in Lemma~\ref{lm:stabStdForm}, because of which the 
primary generators were required to be implemented in a particular order.
If however, we had them strictly in the standard form of Lemma~\ref{lm:stabStdForm}
then any order is possible.  For instance, implementing the generators in the
reverse order for the $[[5,1,3]]$ will give the following circuit.
\begin{figure}[htb]
\[
\Qcircuit @C=.7em @R=.4em @! {
\lstick{\ket{0}}&\qw&\gate{H}& \ctrl{1}&\qw &\qw&\qw& \gate{Z}\qwx[1] & \qw \\
\lstick{\ket{0}}&\qw&\qw&\gate{Z}\qwx[2]& \gate{H} & \ctrl{1}&\qw & \gate{Z}\qwx[1]&\qw&\\
\lstick{\ket{0}}&\qw&\qw& \qw & \qw&\gate{Z}\qwx[1]&\gate{H} & \ctrl{2}& \qw & \\
\lstick{\ket{0}}&\gate{X}& \qw& \gate{X}\qwx[1]&\qw& \gate{X}\qwx[1] &\qw& \qw &\qw \\
\lstick{\ket{\psi}}&\ctrl{-1}&\qw& \gate{Z}&\qw & \gate{Y} & \qw&\gate{Y} &\qw \\
&\ol{X}  \gategroup{4}{2}{5}{2}{.7em}{--}&& G_1  \gategroup{3}{3}{5}{4}{.7em}{_\}}&& G_2  \gategroup{2}{5}{5}{6}{.7em}{_\}}& &G_3 \gategroup{1}{7}{5}{8}{.7em}{_\}}\\
}
\]
\caption{Alternative encoding for the $[[5,1,3]]$ code}
\end{figure}
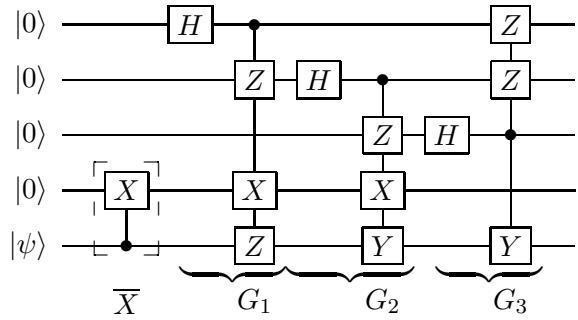
However, note that some additional $Z$ gates are present in this implementation for
$G_1$, while some of the $Z$ gates associated to $G_1$ and $G_2$ are redundant. The 
reduced circuit will be given as:
\begin{figure}[htb]
\[
\Qcircuit @C=.7em @R=.4em @! {
\lstick{\ket{0}} &\qw&\gate{H}& \ctrl{3}&\qw &\qw&\qw& \gate{Z}\qwx[1] & \qw \\
\lstick{\ket{0}} &\qw&\qw&\qw& \gate{H} & \ctrl{2}&\qw & \gate{Z}\qwx[1] & \qw & \\
\lstick{\ket{0}} &\qw&\qw& \qw & \qw&\qw&\gate{H} & \ctrl{2}& \qw & \\
\lstick{\ket{0}}&\gate{X}& \qw& \gate{X}\qwx[1]&\qw& \gate{X}\qwx[1] &\qw& \qw &\qw \\
\lstick{\ket{\psi}}&\ctrl{-1}&\qw& \gate{Z}&\qw & \gate{Y} & \qw&\gate{Y} &\qw \\
&\ol{X}  \gategroup{4}{2}{5}{2}{.7em}{--}&& G_1  \gategroup{3}{3}{5}{4}{.7em}{_\}}&& G_2  \gategroup{2}{5}{5}{6}{.7em}{_\}}& &G_3 \gategroup{1}{7}{5}{8}{.7em}{_\}}\\
}
\]
\caption{Alternative encoding for the $[[5,1,3]]$ code with redundant $Z$ gates removed}
\end{figure}
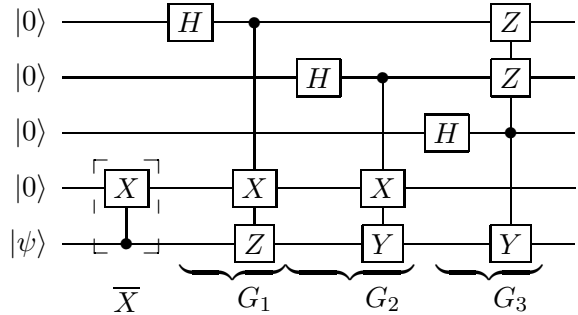

Other orderings of the primary generators are possible, but we must be careful to account
for the $Z$ gates that do not act on the $\ket{0}$ state directly.
\end{remark}

\paragraph{Encoding Clifford Codes.}
In this section, we show that a Clifford code can be encoded using its stabilizer
and therefore the methods used for encoding stabilizer codes are applicable. 
We briefly recapitulate some facts about Clifford subsystem codes.
\index{Clifford code}
Let $E$ be an abstract error group {\em i.e.}, it is a finite group with a faithful 
irreducible unitary representation $\rho$ of degree $|E:Z(E)|^{1/2}$. Denote by $\phi$, 
the irreducible character afforded by $\rho$. Let $N$ be
a normal subgroup of $E$. Further, let $\chi$ be an irreducible character $\chi$ 
of $N$ such that $(\phi_N,\chi)_N >0$.
Then the Clifford code defined by $(E,\rho,N, \chi)$ is the image of the orthogonal 
projector \index{projector!Clifford code}
\begin{eqnarray}
P=\frac{\chi(1)}{|N|}\sum_{n\in N}\chi(n^{-1})\rho(n).\label{eq:cliffProj}
\end{eqnarray}

Under certain conditions we can construct a subsystem code from the Clifford code,
in particular when $E$ is the extraspecial $p$-group, the Clifford code $C$ has a
tensor product decomposition\footnote{Strictly speaking the equality should be
replaced by an isomorphism.} as $C=A\otimes B$, where $B$ is an irreducible 
$\C N$-module, $A$ is an irreducible $\C L$-module and $L=C_E(N)$.
In this case we can encode information only into the subsystem $A$, while the
co-subsystem $B$ provides additional protection. When encoded this
way we say $C$ is a Clifford subsystem code. The normal subgroup $N$ 
consists of all errors in $E$ that act trivially on $A$. It is also 
called the gauge group of the subsystem code. Our main goal will be to show how to encode into 
the subsystem $A$. Therefore,  our interest will center on the projectors
for the Clifford code and the subsystem code and not so much on the 
parameters of the codes themselves. 

An alternate projector for a Clifford code with data $(E,\rho,N,\chi)$ can be defined 
in terms of $Z(N)$, the center of $N$. The proof of this can be found in 
\cite[Theorem~6]{klappenecker033}. This projector is given as \index{projector!Clifford code}
\begin{eqnarray}
P'=\frac{1}{|Z(N)|}\sum_{n\in Z(N)}\varphi(n^{-1})\rho(n),\label{eq:cliffZProj}
\end{eqnarray}
where $\varphi$ is an irreducible character of $Z(N)$, that satisfies 
$(\chi \downarrow Z(N))(x)=\chi(1)\varphi(x)$.  
In this case $Q$ can be thought of as a stabilizer
code in the sense of \cite{calderbank98} i.e. 
\begin{eqnarray}
\rho(m) \ket{\psi} = \varphi(m)\ket{\psi} \mbox{ for any $m$ in } Z(N).
\end{eqnarray}
In addition to the assumption that the error group is an
extraspecial $p$-group we also assume that $Z(E) \le N$.  The inclusion of the center of 
$E$ does not change the code but helps in analysis. Thus we have the following lemma.

\begin{lemma}\label{lm:phivalues}
Let $(E,\rho,N,\chi)$ be the data of a Clifford code and $\varphi$ an irreducible character of $Z(N)$,
the center of $N$, satisfying $(\chi\downarrow Z(N))(x)=\chi(1)\varphi(x)$. If $E$ is an extraspecial $p$-group, then for all $n$ in $Z(N)$,  $\varphi(n) \in\{ \zeta^k \mid \zeta=e^{j2\pi k/p}, 0\leq k<p\}$. Further, 
if $Z(E)\le N$, then for any  $n\in Z(N)$, we have $\varphi(n^{-1}) \rho(n) \in \rho(Z(N))$. 
\end{lemma}
\begin{proof}
First we note that the irreducibilty of $\rho$ implies that for any $z$ in $Z(E)$ we have 
$\rho(z)=  \omega I$ for some $\omega \in \C$ by Schur's lemma. The 
assumption that $E$ is an extraspecial $p$-group
forces $\omega \in \{ \zeta^k \mid 0\leq k<p \}$ where $\zeta=e^{j2\pi/p}$. 
This is because $|Z(E)|=p$ for extraspecial $p$-groups.
Secondly, we observe that $\varphi$ is an irreducible additive character of $Z(N)$ 
(an abelian subgroup of an extraspecial $p$-group) which implies that we must have $\varphi(n) = \zeta^{l}$ for some $0\leq l<p$, \cite{lidl97}.
Together these observations imply that we can assume 
$\varphi(n^{-1}) I= \zeta^{l} I =\rho(z)$  for some $0\leq l\le p$ and
$z\in Z(E)$. Since $Z(E) \le N$, it follows that $Z(E)\le Z(N)$ and $\varphi(n^{-1})\rho(n)$ 
is in $\rho(Z(N))$.
\end{proof}

Our goal is to use the stabilizer of $Q$ for encoding and as a first step
we will show that it can be computed from $Z(N)$.
The usefulness of such a projector is
that it obviates the need to know the character $\varphi$. Let $S \le \rho(E)$ be the
stabilizer of $Q$. Then we claim that $S$ is given as 
$$
S= \{\varphi(n^{-1})\rho(n) \mid n \in Z(N) \}.
$$
We claim that $S$ can be used for encoding the 
associated Clifford code. Then we will show how the encoding circuit of the
Clifford code is to be modified so that
we can encode the subsystem code derived from the Clifford code. 
\begin{theorem}\label{th:projClifford}
Let $Q$ be a Clifford code with the data $(E,\rho,N,\chi)$ and
$\varphi$ a constituent of the restriction of $\chi$ to $Z=Z(N)$. 
Let $E$ be an extraspecial $p$-group and $Z(E)\le N$ and 
\begin{eqnarray}
S=\left\{ \varphi(n^{-1}) \rho(n) \mid n\in Z(N) \right\} \quad \mbox{and}\quad
P=\frac{1}{|S|}\sum_{s\in S} s .\label{eq:stabProj}
\end{eqnarray}
Then $S$ is the stabilizer of $Q$ and $\text{Im } P =Q$.
\end{theorem}
\begin{proof}
We will show this in a series of steps.
\begin{enumerate}[1)]
\item
First we will show that $S\leq \rho(Z)$. By Lemma~\ref{lm:phivalues} we know that
$\varphi(n^{-1})\rho(n)$ is in $\rho(Z)$, therefore $S\subseteq \rho(Z)$. For any two
elements $n_1,n_2\in Z$, we have $s_1=\varphi(n_1^{-1})\rho(n_1), s_2=\varphi(n_2^{-1})\rho(n_2) \in S$ and we can verify that  $s_1^{-1}s_2 = \varphi(n_1)\rho(n_1^{-1})\varphi(n_2^{-1})\rho(n_2)=
\varphi(n_2^{-1}n_1)\rho(n_1^{-1}n_2) \in S$, as  $\rho(n_1^{-1}n_2)$ is in $\rho(Z)$. 
Hence $S \leq \rho(Z)$.
\item Now we show that $S$ fixes $Q$. Let $s\in S$ and $\ket{\psi}\in Q$. Then
$s=\varphi(n^{-1})\rho(n)$ for some $n\in Z$. The action of $s$ on $\ket{\psi}$
is given as $s\ket{\psi} =\varphi(n^{-1})\rho(n)\ket{\psi}= \varphi(n^{-1})\varphi(n)\ket{\psi}=\ket{\psi}$, in other words $S$ fixes $Q$. 
\item  Next, we show that $|S|=|Z|/|Z(E)|$.
If two elements $n_1$ and $n_2$ in $Z$ map to the same element in $S$, then 
$\varphi(n_1^{-1})\rho(n_1) = \varphi(n_2^{-1})\rho(n_2)$, that is 
$\rho(n_2) = \varphi(n_1^{-1}n_2)\rho(n_1)$. From Lemma~\ref{lm:phivalues} it follows that
$\rho(n_2)=\zeta^l \rho(n_1)$ for some $0\leq l<p$. 
Since $\rho(Z(E))=\{e^{j2\pi k/p } I\mid 0\leq k<p \}$, we must have $n_2=zn_1$ for some $z\in Z(E)$. 
Thus, $|S|=|Z|/|Z(E)|$.
\item
Let $T$ be a traversal of  $Z(E)$ in $Z$, then every element in $Z$
can be written as $zt$ for some $z\in Z(E)$ and $t\in T$. From step 3)
we can see that all elements in a coset of $Z(E)$ in $Z$ map to the same element
in $S$, therefore, 
$$S=\{\varphi(t^{-1})\rho(t) \mid t\in T \}.$$
Recall that a projector for $Q$ is given by  
\begin{eqnarray*}
P' &=& \frac{1}{|Z|}\sum_{n\in Z}\varphi(n^{-1})\rho(n) ,\\
&=& \frac{1}{|Z|}\sum_{t\in T}\sum_{z\in  Z(E)}\varphi((zt)^{-1})\rho(zt).
\end{eqnarray*}
But we know from step 3) that if $z\in Z(E)$, then $\varphi(n^{-1})\rho(n)  = \varphi((zn)^{-1})\rho(z n)$. 
So we can simplify $P'$ as 
\begin{eqnarray*}
P'&=&\frac{1}{|Z|}\sum_{t\in T}\sum_{z\in  Z(E)}\varphi(t^{-1})\rho(t),\\
&=&\frac{|Z(E)|}{|Z|}\sum_{t\in T}\varphi(t^{-1})\rho(t)\\
&=&\frac{1}{|S|}\sum_{s\in S} s = P.
\end{eqnarray*}
Thus the projector defined by $S$ is precisely the same as $P'$
and $P$ is also a projector for $Q$. 
%
\end{enumerate}
From step 3) it is clear that $S\cap Z(E) =\{ \mathbf{1}\}$ and by 
\cite[Lemma~10]{ketkar06}, 
$S$ is a closed subgroup of $E$. By
\cite[Lemma~9]{ketkar06}, 
$\text{Im }P =Q$ is a stabilizer code. 
Hence $S$ is the stabilizer of $Q$.
\end{proof}

\begin{corollary}\label{co:projSubsys}
Let $Q$ be an $[[n,k,r,d]]$ Clifford subsystem code and $S$ its stabilizer. Let
\begin{eqnarray}
P = \frac{1}{|S|}\sum_{s\in S} s.
\end{eqnarray}
Then $P$ is a projector for the subsystem code {\i.e.} $Q = \text{Im } P$. 
\end{corollary}
\begin{proof}
By \cite[Theorem~4]{pre06}, we know that an $[[n,k,r,d]]$ Clifford subsystem code is derived from a
Clifford code with data $(E,\rho,N,\chi)$. This construction assumes that 
$E$ is an extraspecial $p$-group and $Z(E)\le N  \trianglelefteq E$. 
Since as subspaces the Clifford code and subsystem code are identical, 
by Theorem~\ref{th:projClifford} we conclude that
the projector defined from the stabilizer of the subspace is also a projector for the 
subsystem code. 
\end{proof}

Theorem~\ref{th:projClifford} shows that any Clifford code can be encoded using its
stabilizer. 
As to a subsystem code, while Corollary~\ref{co:projSubsys} shows that 
there exists a projector that can be defined from its stabilizer, it is not clear 
how to use it so that one respects the subsystem structure during encoding.
More precisely, how do we use the projector defined in Corollary~\ref{co:projSubsys} to encode
into the information carrying subsystem $A$ and not the gauge subsystem. This will
be the focus of the next section.

\paragraph{Encoding Subsystem Codes.}

For ease of  presentation and clarity henceforth we will focus on binary codes, 
though the results can be  extended to nonbinary alphabet using  methods 
similar to stabilizer codes, see \cite{grassl03}. 
Theorem~\ref{th:projClifford} shows that in order to encode Clifford codes we can use 
a projector derived from the underlying stabilizer to project onto the codespace. 
But in case of Clifford subsystem codes 
we know that $Q=A\otimes B$ and the information is to be actually encoded in $A$.
Hence, it is not sufficient to merely project onto $Q$, we must also show that 
we encode into $A$ when we encode using the projector defined in 
Corollary~\ref{co:projSubsys}.

Let us clarify what we mean by encoding the information in $A$
and not in $B$. Suppose that $P$ maps $\ket{0}$  to $\ket{\psi}_A\otimes \ket{0}_B$
and $\ket{1}$ to $\ket{\psi}_A\otimes \ket{1}_B$. Then the
information is actually encoded into $B$. Since the gauge group acts nontrivially on $B$,
this particular encoding does not protect information. 
Of course a subsystem code should not encode (only) into $B$, but we have to show that the 
projector defined by $P_s$ does not do that.

We need the following result on the structure of the gauge group and the encoded operators
of a subsystem code. Poulin \cite{poulin05} proved a useful result on the structure of the
gauge group and the encoded operators of the subsystem code. 
But first a little notation.
A basis for $\mathcal{P}_n$ is $X_i,Z_i$, $1\leq i\leq n$, where $X_i$ and $Z_i$
are given as 
$$
X_i =\bigotimes_{j=1}^n X^{\delta_{ij}} \quad \mbox{ and } \quad Z_i =\bigotimes_{j=1}^n Z^{\delta_{ij}}.
$$
They satisfy the relations  $[X_i,X_j]=0=[Z_i,Z_j]$; $[X_i,Z_j]=2\delta_{ij}X_iZ_j$. However,
we can choose other generating sets $\{x_i,z_i\mid 1\leq i \leq n \}$ for $\mathcal{P}_n$
that satisfy similar commutation relations {\em i.e.}, 
$[x_i,x_j]=0=[z_i,z_j]$ and $[x_i,z_j]=2\delta_{ij}x_iz_j$. These operators
may act nontrivially on many qubits. 
We often refer to the pair of operators $x_i,z_i$ that satisfy the commutation relations
similar to the Pauli operators as a \textsl{hyperbolic pair}. 
Given an $[[n,k,r,d]]$ code we could view the state space of the physical 
$n$ qubits as that of $n$ virtual qubits on which these $x_i,z_i$ act as 
$X$ and $Z$ operators. In particular $k$ of these virtual qubits are the 
logical qubits and $r$ of them  gauge qubits. The usefulness of these
operators is that we can specify the structure of the stabilizer, the
gauge group and the encoded operators. The following lemma makes
this specification precise. 
\begin{lemma}\label{lm:struct}
Let $Q$ be an $[[n,k,r,d]]_2$ subsystem code with gauge group, $G$ and stabilizer $S$.
Denote the encoded operators by $\ol{X}_i,\ol{Z}_i$, $1\leq i\leq k$, where 
$[\ol{X}_i,\ol{X}_j]=0=[\ol{Z}_i,\ol{Z}_j]; [\ol{X}_i,\ol{Z}_j ]=2  \delta_{ij}\ol{X}_i \ol{Z}_j$. Then 
there exist operators $\{ x_i,z_i \in \mathcal{P}_n \mid 1\leq i\leq n\}$ such that 
\begin{compactenum}[i)]
\item $S= \langle z_1, z_2, \ldots, z_s \rangle$, 
\item $G = \langle S, z_{s+1},x_{s+1},\ldots, z_{s+r},x_{s+r}, Z({\mathcal{P}_n}) \rangle$, 
\item $C_{\mathcal{P}_n}(S) = \langle G, \ol{X}_1,\ol{Z}_1,\ldots,  \ldots, \ol{X}_k, \ol{Z}_k \rangle$,
\item $\ol{X}_i =x_{s+r+i}$ and $\ol{Z}_i=z_{s+r+i}$, $1\leq i\leq k$,
\end{compactenum}
where 
$[z_i,z_j]=[x_i,x_j]=0; [x_i,z_i]=2\delta_{ij}x_iz_i$. Further, $S$ defines an $[[n,k+r]]$ stabilizer
code encoding into the same space as the subsystem code and its encoded operators are given
by $\{x_{s+1},z_{s+1},\ldots, x_{s+r},z_{s+r},  \ol{X}_1,\ol{Z}_1, \ldots, \ol{X}_k,\ol{Z}_k\}$
\end{lemma}
\begin{proof}
See \cite{poulin05} for proof on the structure of the groups.
Let $Q=A\otimes B$, then $\dim A =2^k$ and $\dim B= 2^r$.
From Corollary~\ref{co:projSubsys} we know that the projector defined by $S$  also 
projects onto $Q$ (which is $2^{k+r}$-dimensional) and therefore it defines an 
$[[n,k+r]]$ stabilizer code. 
From the definition of the operators $x_i,z_i$  and  $\ol{X}_{i}, \ol{Z}_i$ and the fact that
$$C_{\mc{P}_n}(S) = \langle S,x_{s+1},z_{s+1},\ldots, x_{s+r},z_{s+r} \ol{X}_1,\ol{Z}_1,\ldots, \ol{X}_k,\ol{Z}_k, Z({\mathcal{P}_n}) \rangle $$ 
we see that 
$x_i,z_i$, for $ s+1\leq i\leq r$ act like encoded operators on the gauge qubits, 
while  $\ol{X}_{i}, \ol{Z}_i$ continue to be the encoded operators on the information qubits.
Together they exhaust the set of $2(k+r)$ encoded operators of the $[[n,k+r]]$ stabilizer code. 
\end{proof}

We observe that the logical operators of the subsystem code are also logical operators for the
underlying stabilizer code. So if the 
stabilizer code and the subsystem code have the same logical all zero state, then 
Lemma~\ref{lm:struct} suggests that in order to encode the subsystem code, we can treat it as
stabilizer code and use the same techniques to encode. If the logical all zero code word was the same for
both the codes, then  because they have the same logical operators we can encode any given 
input to the same logical state in both cases. Using linearity we could then encode any
arbitrary state. Encoding  the all zero state seems to be the key. Now, even in the case
of the stabilizer codes, there is no unique all zero logical state. There are many possible
choices. The reader can refer to the appendix for examples. Given the 
encoded operators it is easy to define the logical all zero state as the following definition
shows: 

\begin{defn}\label{def:logZero}
A logical all zero state of an $[[n,k,r,d]]$ subsystem code is any state that is
fixed by its stabilizer and $k$ logical $Z$ operators. 
\end{defn}
This definition is valid in case of stabilizer codes also. This definition might appear a little circular. After all, we seem to have assumed the definition of the logical $Z$ operators. Actually, 
this is a legitimate definition because, depending on the choice
of our logical operators, we can have many choices of the  logical all zero state. In case of the
subsystem codes, this definition implies that the logical all zero state is fixed by $n-r$ operators,
consequently it can be any state in that $2^r$-dimensional subspace. If we consider the $[[n,k+r]]$
stabilizer code that is associated to the subsystem code, then its logical zero is additionally
fixed by $r$ more operators. So any logical zero of the stabilizer code is also a logical all zero  state
of the subsystem code. It follows that if we know how to encode the stabilizer code's logical all zero,
we know how to encode the subsystem code. We are
interested in more than merely encoding the subsystem code of course. We also want  to leverage
the gauge qubits to simplify and/or make the encoding process more robust. 
Perhaps a few examples will clarify the ideas. 

\begin{figure}[ht]
\[
\Qcircuit @C=.7em @R=.4em @! {
\lstick{\ket{0}} & \qw & \gate{H}& \ctrl{3} & \qw \\
\lstick{\ket{g}} & \qw & \ctrl{2} & \targ & \qw & \\
\lstick{\ket{\psi}} & \ctrl{1} & \qw & \targ & \qw & \\
\lstick{\ket{0}} & \targ & \targ & \targ & \qw & \\
}
\]
\caption{Encoding the $[[4,1,1,2]]$ code (Gauge qubits can be in any state)}\label{fig:stdFormEncEx1}
\end{figure}
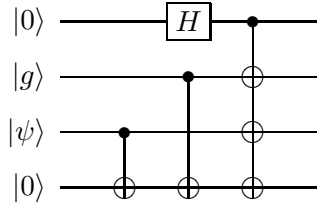

\paragraph{Illustrative Examples.}
Consider the following $[[4,1,1,2]]_2$ subsystem code, with the gauge group $G$,
stabilizer $S$ and encoded operators given by $L$.
\begin{eqnarray*}
S&=&\left[ \begin{array}{cccc}X&X&X&X\\Z&Z&Z&Z 
\end{array}\right] =\left[ \begin{array}{c}z_1\\z_2\end{array}\right],\\
G&=&\left[ \begin{array}{cccc} 
X&X&X&X\\Z&Z&Z&Z \\  \hline
I&X&I&X \\I&I&Z&Z
\end{array}\right] = \left[ \begin{array}{c}  z_1\\z_2\\\hline x_3\\z_3\end{array}\right].
\end{eqnarray*} 
The encoded operators of this code are given by 
\begin{eqnarray*}
L&=&\left[ \begin{array}{cccc} I&I&X&X \\I&Z&I&Z\end{array}\right]=\left[ \begin{array}{c}\ol{X}_1\\\ol{Z}_1\end{array}\right].
\end{eqnarray*} 
The associated $[[4,2]]$ stabilizer code has the following encoded operators.
\begin{eqnarray*}
T&=&\left[ \begin{array}{cccc}  I&X&I&X \\ I&I&X&X\\
 I&I&Z&Z\\ I&Z&I&Z 
\end{array}\right]=\left[ \begin{array}{c}x_3\\\ol{X}_1 \\
z_3 \\\ol{Z}_1 \end{array}\right].
\end{eqnarray*} 
It will be observed that the encoded $X$ operators of $[[4,2]]$ are in a
form convenient for encoding. 
We treat the $[[4,1,1,2]]$ code as $[[4,2]]$ code and encode it as in 
Figure~\ref{fig:stdFormEncEx1}. The gauge qubits are permitted to be 
in any state. 

Assuming $g=a\ket{0}+b\ket{1}$, the logical states up to a normalizing constant are
\begin{eqnarray*}
\ket{\ol{0}} &=&a(\ket{0000}+\ket{1111})+b(\ket{0101}+\ket{1010}),\\
\ket{\ol{1}} &=&a(\ket{0011}+\ket{1100})+b(\ket{0110}+\ket{1001}).
\end{eqnarray*}
It can be easily verified that $S$ stabilizes the above state and while 
the gauge group acts in a nontrivial fashion, the resulting states are still
orthogonal. 
In this example we have encoded as if we were encoding the $[[4,2]]$ code. Prior to 
encoding the gauge qubits can be identified with physical qubits. After the encoding
however such a correspondence between the physical qubits and gauge qubits does not 
necessarily exist in a nontrivial subsystem code. Since the encoded operators
of the subsystem code are also encoded operators for the stabilizer code, we are
guaranteed that the information is not encoded into the gauge subsystem.

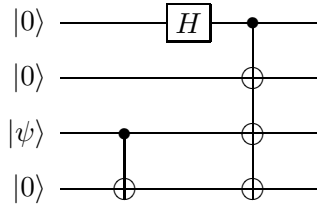
\begin{figure}[ht]
\[
\Qcircuit @C=.7em @R=.4em @! {
\lstick{\ket{0}} & \qw & \gate{H}& \ctrl{3} & \qw \\
\lstick{\ket{0}} & \qw & \qw & \targ & \qw & \\
\lstick{\ket{\psi}} & \ctrl{1} & \qw & \targ & \qw & \\
\lstick{\ket{0}} & \targ & \qw & \targ & \qw & \\
}
\]
\caption{Encoding the $[[4,1,1,2]]$ code (Gauge qubits initialized to zero)}\label{fig:stdFormEncEx1Opt}
\end{figure}

As the state of gauge qubits is of no consequence, we can initialize them to any state.
Alternatively, if we initialized them to zero, we can simplify the circuit as shown in 
Figure~\ref{fig:stdFormEncEx1Opt}.

The encoded states in this case are (again, the normalization factors are ignored)
\begin{eqnarray*}
\ket{\ol{0}} &=&\ket{0000}+\ket{1111},\\
\ket{\ol{1}} &=&\ket{0011}+\ket{1100}.
\end{eqnarray*}
The benefit with respect to the previous version is that at the cost of initializing
the gauge qubits, we have been able to get rid of all the encoded operators associated
with them. This seems to be a better option than randomly initializing the gauge qubits.
Because it is certainly easier to prepare them in a known state like $\ket{0}$, rather
than implement a series of controlled gates depending on the encoded operators associated
with those qubits. 

At this point we might ask if it is possible to get both the benefits of random initialization
of the gauge qubits as well as avoid implementing the encoded operators associated with them.
To answer this question let us look a little more closely at the previous two encoding circuits for the
subsystem codes. We can see from them that it will not work in general. Let us see why. If 
we initialize the gauge qubit to $\ket{1}$ instead of $\ket{0}$ in the encoding given in 
Figure~\ref{fig:stdFormEncEx1Opt}, then  the  encoded state is 
\begin{eqnarray*}
\ket{\ol{0}} &=&\ket{0100}+\ket{1011},\\
\ket{\ol{1}} &=&\ket{0111}+\ket{1000}.
\end{eqnarray*}
Both these states are not stabilized by $S$, indicating that these states are not in the
code space. 

In general, an encoding circuit where it is simultaneously possible 
initialize the gauge qubits to random states and also avoid the encoded operators is likely
to be having more complex primary generators. For instance, let us consider the 
following $[[4,1,1,2]]$ subsystem code:
\begin{eqnarray*}
S&=&\left[ \begin{array}{cccc}X&Z&Z&X\\Z&X&X&Z 
\end{array}\right] =\left[ \begin{array}{c}z_1\\z_2\end{array}\right],\\
G&=&\left[ \begin{array}{cccc} 
X&Z&Z&X\\Z&X&X&Z \\  \hline
Z&I&X&I \\I&Z&Z&I
\end{array}\right] = \left[ \begin{array}{c}  z_1\\z_2\\\hline x_3\\z_3\end{array}\right].
\end{eqnarray*} 
The encoded operators of this code are given by 
\begin{eqnarray*}
L&=&\left[ \begin{array}{cccc} I&Z&I&X \\Z&I&I&Z\end{array}\right]=\left[ \begin{array}{c}\ol{X}_1\\\ol{Z}_1\end{array}\right].
\end{eqnarray*} 
The associated $[[4,2]]$ stabilizer code has the following encoded operators.
\begin{eqnarray*}
T&=&\left[ \begin{array}{cccc}  Z&I&X&I \\
I&Z&I&X \\
I&Z&Z&I\\
Z&I&I&Z
\end{array}\right]=\left[ \begin{array}{c}x_3\\\ol{X}_1 \\
z_3 \\\ol{Z}_1 \end{array}\right].
\end{eqnarray*} 
The encoding circuit for this code is given by 
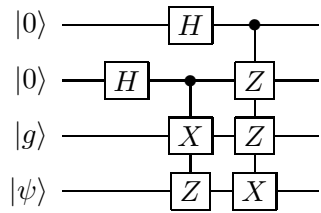
\begin{figure}[htb]
\[
\Qcircuit @C=.7em @R=.4em @! {
\lstick{\ket{0}} & \qw & \gate{H}& \ctrl{1} & \qw \\
\lstick{\ket{0}} & \gate{H} & \ctrl{1} & \gate{Z}\qwx[1] & \qw & \\
\lstick{\ket{g}} & \qw & \gate{X}\qwx[1] & \gate{Z} \qwx[1]& \qw & \\
\lstick{\ket{\psi}} & \qw & \gate{Z} & \gate{X} &\qw \\
}
\]
\caption{Encoding $[[4,1,1,2]]$ code (Encoded operators for the gauge qubits are trivial and 
gauge qubits can be initialized to random states)}
\end{figure}


In this particular case, the gauge qubits (as well as the information qubits)
do not require any additional encoding circuitry. In this case we can 
initialize the gauge qubits to any state we want.
But, the reader would have observed we did not altogether end up with 
a simpler circuit. The primary generators are two as against one and the
complexity of the encoded operators has been shifted to them.
So even though we were able to  get rid of the encoded operator on the gauge qubit 
and also get the benefit of initializing it to a random state, this is still more 
complex compared to either of encoders in Figures~\ref{fig:stdFormEncEx1} and \ref{fig:stdFormEncEx1Opt}.
 Our contention is that it is better to initialize
the gague qubits to zero state and not implement the encoded operators associated
to them. 

\paragraph{Encoding Subsystem Codes by Standard Form Method.}
The previous two examples might lead us to conclude that we can take the stabilizer
of the given subsystem code and form the encoded operators by reducing the stablizer
to its standard form and encode as if it were a stabilizer code. However, there are certain
subtle points to be kept in mind.
When we form the encoded operators we get $k+r$
encoded operators; we cannot from the stabilizer alone conclude which are the encoded
operators on the information qubits and which on the gauge qubits. Put differently,
these operators belong to the space $C_{\mc{P}_n}(S)\setminus S = G C_{\mc{P}_n}(G) \setminus S Z(\mc{P}_n)$. It is not guaranteed that they are entirely in $C_{\mc{P}_n}(G)$ {\em i.e.}, we cannot say if they act as encoded operators on the logical qubits. This implies that in general all these operators act nontrivially
on both $A$ and $B$. Consequently, we must be careful in choosing the encoded operators and
the gauge group must be taken into account. We give two slightly different methods for 
encoding subsystem codes. The difference between the two methods is subtle. Both methods
require the gauge qubits to be initialized to zero. In the second method (see Algorithm~\ref{alg:subsysEncOpt}) however,
we can avoid the encoded operators associated to them. Under certain circumstances, we can also 
permit initialization to random states.

\begin{algorithm}
\caption{{\ensuremath{\mbox{\scshape Encoding subsystem codes -- Standard form method 1}}}
}\label{alg:subsysEnc}
\begin{algorithmic}[1]
\REQUIRE Gauge group, $G=\langle S, x_{s+1},z_{s+1},\ldots, x_{s+r},z_{s+r}, \pm I \rangle$
 and stabilizer, $S =\langle z_1,\ldots, z_{n-k-r}\rangle$ of the $[[n,k,r,d]]$ subsystem code.

\ENSURE $[x_i,x_j] =[z_i,z_j] =0$; $[x_i,z_j ]=2x_iz_i \delta_{ij}$
\medskip
\STATE Form $S_A= \langle S, z_{s+1}, \ldots, z_{s+r}\rangle $, where $s=n-k-r$
\STATE Compute the standard form of $S_A$ as per Lemma~\ref{lm:stabStdForm}
$$S_A =_{\pi} \left[\begin{array}{ccc|ccc}
I_{s'}& A_1 & A_2 & B & 0 & C\\
0 & 0 & 0& D&I_{s+r-s'} & E
\end{array} \right] $$

\STATE Compute the encoded operators $\ol{X}_1,\ldots, \ol{X}_k$ as
$$
\left[\begin{array}{c}
\ol{Z}\\
\ol{X} 
\end{array}\right]
=_{\pi}\left[\begin{array}{ccc|ccc}
0 & 0 &0 &A_2^t&0 &I_{k}\\
\hline
0 & E^t &I_{k}&C^t&0 &0 
\end{array}\right]
$$
\STATE Encode using the primary generators of $S_A$ and $\ol{X}_i$ as encoded operators, 
see Lemma~\ref{lm:implement}; all the other $(n-k)$ qubits are initialized to $\ket{0}$.
\end{algorithmic}
\end{algorithm}

\textbf{Correctness of Algorithm~\ref{alg:subsysEnc}.}
Since stabilizer $S_A \ge S$, the space stabilized by $S_A$ is  a subspace of the $A\otimes B$,
the subspace stabilized by $S$. As $|S_A|/|S|=2^r$, the dimension of the subspace stabilized by $S_A$ is
$2^{k+r}/2^r=2^k$. 
Additionally, the generators $z_{s+1},\ldots, z_{s+r}$ act trivially on 
$A$. The encoded operators as computed in the algorithm act nontrivially on $A$ and give 
$2^k$ orthogonal states; thus we are assured that the information is encoded into $A$. 

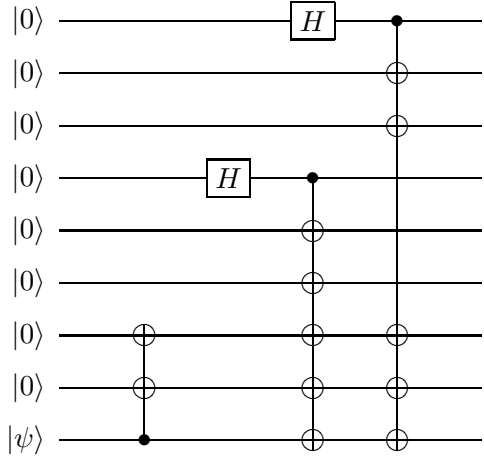
\begin{figure}[htb]
\[
\Qcircuit @C=1.4em @R=0.3em @!{
\lstick{\ket{0}}&\qw&\qw&\gate{H} &\ctrl{8}&\qw\\
\lstick{\ket{0}}&\qw&\qw&\qw &\targ&\qw\\
\lstick{\ket{0}}&\qw&\qw&\qw &\targ&\qw\\
\lstick{\ket{0}}&\qw&\gate{H} &\ctrl{5}&\qw&\qw\\
\lstick{\ket{0}}&\qw&\qw &\targ&\qw&\qw\\
\lstick{\ket{0}}&\qw&\qw &\targ&\qw&\qw\\
\lstick{\ket{0}}&\targ&\qw &\targ&\targ&\qw\\
\lstick{\ket{0}}&\targ&\qw&\targ&\targ&\qw\\
\lstick{\ket{\psi}}&\ctrl{-2}&\qw&\targ&\targ&\qw\\
}
\]
\caption{Encoder for the $[[9,1,4,3]]$ code. This is also an encoder for the
$[[9,1,3]]$ code.}\label{fig:shorCodeEnc}
\end{figure}

Let us encode the $[[9,1,4,3]]$ Bacon-Shor code using the method just proposed. The stabilizer
and the gauge group are given\footnote{We do not show the identity.} by  
\begin{eqnarray*}
S &= &\left[ \begin{array}{ccc|ccc|ccc}
X&X&X& & & &X&X&X\\
 & & &X&X&X&X&X&X\\
Z& &Z&Z& &Z&Z& &Z\\
 &Z&Z& &Z&Z& &Z&Z
\end{array}\right],
\end{eqnarray*}
\begin{eqnarray*}
G &= &\left[ \begin{array}{ccc|ccc|ccc}
X&X&X& & & &X&X&X\\
 & & &X&X&X&X&X&X\\
Z& &Z&Z& &Z&Z& &Z\\
 &Z&Z& &Z&Z& &Z&Z\\ \hline 
&X&& & & &&X&\\
&&X& & & &&&X\\
 & & &&X&&&X&\\
 & & &&&X&&&X\\\hline 
Z& &Z& & & & & & \\
 & & &Z& &Z& & & \\
 &Z&Z& & & & & & \\
 & & & &Z&Z&&&
 \end{array}\right]
 =\left[ \begin{array}{c}
S\\\hline G_x\\ \hline G_z\end{array}\right].
\end{eqnarray*}
Let us form $S_A$ by augmenting  $S$ with $G_z$. Then
\begin{eqnarray*}
S_A&=&\left[ \begin{array}{ccc|ccc|ccc}
X&X&X& & & &X&X&X\\
 & & &X&X&X&X&X&X\\
Z& &Z&Z& &Z&Z& &Z\\
 &Z&Z& &Z&Z& &Z&Z\\ \hline 
Z& &Z& & & & & & \\
 & & &Z& &Z& & & \\
 &Z&Z& & & & & & \\
 & & & &Z&Z& & & 
\end{array}\right].
\end{eqnarray*}
The encoded $X$ and $Z$ operators are $X_7X_8X_9$ and $Z_1Z_4Z_7$, respectively. 
After putting $S_A$ in the standard form, and encoder for this code is given 
in Figure~\ref{fig:shorCodeEnc}.

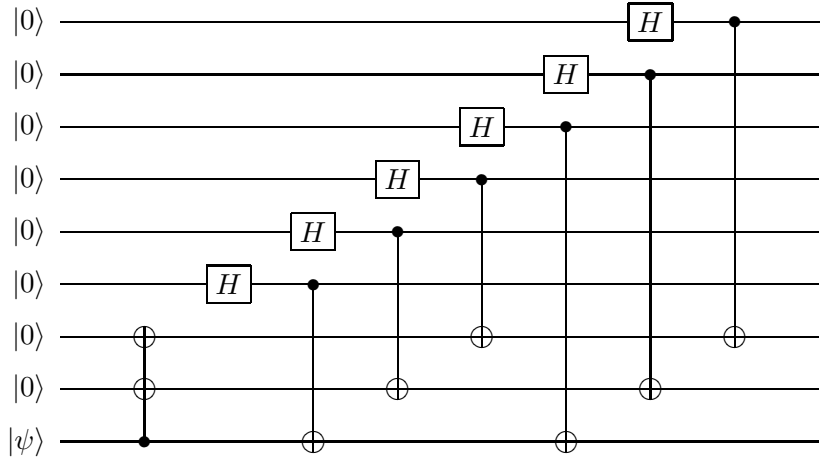
\begin{figure}[h]
\[
\Qcircuit @C=1.4em @R=0.3em @!{
\lstick{\ket{0}}&\qw&\qw&\qw&\qw&\qw &\qw&\gate{H}&\ctrl{6}&\qw\\
\lstick{\ket{0}}&\qw&\qw&\qw&\qw&\qw &\gate{H}&\ctrl{6}&\qw&\qw\\
\lstick{\ket{0}}&\qw&\qw&\qw&\qw&\gate{H}&\ctrl{6}&\qw&\qw&\qw\\
\lstick{\ket{0}}&\qw&\qw&\qw&\gate{H} &\ctrl{3}&\qw&\qw&\qw&\qw\\
\lstick{\ket{0}}&\qw&\qw &\gate{H}&\ctrl{3}&\qw&\qw&\qw&\qw&\qw\\
\lstick{\ket{0}}&\qw&\gate{H} &\ctrl{3}&\qw&\qw&\qw&\qw&\qw&\qw\\
\lstick{\ket{0}}&\targ&\qw &\qw&\qw&\targ&\qw&\qw&\targ&\qw\\
\lstick{\ket{0}}&\targ&\qw&\qw&\targ&\qw&\qw&\targ&\qw&\qw\\
\lstick{\ket{\psi}}&\ctrl{-2}&\qw&\targ&\qw&\qw&\targ&\qw&\qw&\qw\\
}
\]
\caption{Encoder for the $[[9,1,4,3]]$ code with fewer CNOT gates.}\label{fig:shorCodeEncFewer}
\end{figure}

If on the other hand we had formed $S_A$ by adding $G_x$ instead, then 
$S_A$ would have been 
\begin{eqnarray*}
S_A=\left[ \begin{array}{ccc|ccc|ccc}
X& & & & & &X& & \\
 &X& & && & &X& \\
 & &X& & && & &X\\
 & & &X& & &X& & \\
 & & & &X& & &X& \\
 & & & & &X& & &X\\
Z& &Z&Z& &Z&Z& &Z\\
 &Z&Z& &Z&Z& &Z&Z\\ 
\end{array}\right].
\end{eqnarray*}
The encoded operators remain the same. In this case the encoding circuit is given 
in Figure~\ref{fig:shorCodeEncFewer}.

The circuit in Figure~\ref{fig:shorCodeEncFewer} has fewer CNOT gates, though the number of single qubit gates has
increased. Since we expect the implementation of the CNOT gate to be
more complex than the $H$ gate, this might be a better choice. In any case,
this demonstrates that by exploiting the gauge qubits one can find ways
to reduce the complexity of encoding circuit.

The gauge qubits provide a great degree of freedom in encoding.  We consider the following
variant on standard form encoding, where we try to minimize the 
the number of primary generators. This is not guaranteed to reduce
the overall complexity, since that is determined by both the primary generators
and the encoded operators. Fewer primary generators might usually imply encoded operators 
with larger complexity. In fact we have already seen, that in the case of 
$[[9,1,4,3]]_2$ code that a larger number of primary generators does not 
necessarily imply higher complexity.
However, it has the potential for lower complexity.

\begin{algorithm}[H]
\caption{{\ensuremath{\mbox{\scshape Encoding subsystem codes -- Standard form method 2}}}}\label{alg:subsysEncOpt}
\begin{algorithmic}[1]
\REQUIRE Gauge group, $G=\langle S, x_{s+1},z_{s+1},\ldots, x_{s+r},z_{s+r}, \pm I \rangle$
 and stabilizer, $S =\langle z_1,\ldots, z_{n-k-r}\rangle$ of the $[[n,k,r,d]]$ subsystem code.

\ENSURE $[x_i,x_j] =[z_i,z_j] =0$; $[x_i,z_j ]=2x_iz_i \delta_{ij}$
\medskip
\STATE Compute the standard form of $S$ as per Lemma~\ref{lm:stabStdForm}
$$S =_{\pi_1} \left[\begin{array}{ccc|ccc}
I_{s'}& A_1 & A_2 & B & 0 & C\\
0 & 0 & 0& D&I_{s-s'} & E
\end{array} \right] $$

\STATE Form $S_A= \langle S, z_{s+1}, \ldots, z_{s+r}\rangle $, where $s=n-k-r$
\STATE Compute the standard form of $S_A$ as per Lemma~\ref{lm:stabStdForm}
$$S_A =_{\pi_2} \left[\begin{array}{ccc|ccc}
I_{l}& F_1 & F_2 & G_1 & 0 &G_2 \\
0 & 0 & 0& D'&I_{s+r-l} & H
\end{array} \right] $$

\STATE Compute the encoded operators $\ol{X}_1,\ldots, \ol{X}_k$ as
$$
\left[\begin{array}{c}
\ol{Z}\\
\ol{X} 
\end{array}\right]
=_{\pi_2}\left[\begin{array}{ccc|ccc}
0 & 0 &0 &F_2^t&0 &I_{k}\\
\hline
0 & H^t &I_{k}&G_2^t&0 &0 
\end{array}\right]
$$

\STATE Encode using the primary generators of $S$ and $\ol{X}_i$ as encoded operators,
accounting for $\pi_1$ and $\pi_2$, see Lemma~\ref{lm:implement}; all the other $(n-k)$ qubits are initialized to $\ket{0}$.
\end{algorithmic}
\end{algorithm}

The main difference in the second method comes in lines 1 and 5. We encode using the 
primary generators of the stabilizer of the subsystem code instead of the augmented
stabilizer. The encoded operators however remain the same as before. 

\textbf{Correctness of Algorithm~\ref{alg:subsysEncOpt}.}
The correctness of this method lies in the observation we made earlier 
(see discussion following Definition~\ref{def:logZero}), that any logical all zero
state of the stabilizer code is also a logical all zero of the subsystem code and the fact that both share the encoded operators on the encoded qubits. 

\begin{remark}
The permutation $\pi_2$ in Algorithm~\ref{alg:subsysEncOpt} can be restricted to the
last $n-s'$ columns, since while adjoining the additional $r$ generators to $S$,
we could take it to be in the standard form. 
\end{remark}

The encoded operators are given modulo the elements of
the gauge group as in Algorithm~\ref{alg:subsysEnc}, which implies that the their action might be nontrivial on the gauge qubits. The benefit of the second method is when $S$ and $S_A$ have different number of 
primary generators. 
The following aspects of both the methods are worth highlighting. 
\begin{compactenum}[1)]
\item The gauge qubits must be initialized to $\ket{0}$ in both methods. 
\item In Algorithm~\ref{alg:subsysEnc},  the number of primary generators of $S$ 
and $S_A$ can be different leading to a potential increase in complexity compared to 
encoding with $S$.
\item In both methods, the encoded operators as computed are modulo $S_A$. Consequently, 
the  encoded operators might act nontrivially on the gauge qubits. 
\end{compactenum}

\paragraph{Encoding Subsystem Codes by Conjugation Method.} 
The other benefit of subsystem codes is
the random initialization of the gauge qubits.  We now give circuits where we can encode 
the subsystem codes to realize this benefit. But instead of using the standard form method
we will use the conjugation method proposed by Grassl {\em et al.}, \cite{grassl03}
for stabilizer codes. After
briefly reviewing this method we shall show how it can be modified for encoding subsystem
codes. 

The conjugation encoding method can be understood as follows.
It is based on the idea that the Clifford group acts transitively on the Pauli error group. 
It is possible to transform the stabilizer matrix of any $[[n,k,d]]$ stabilizer code 
into the matrix $(0 0 | I_{n-k} 0 )$. For a code with this stabilizer matrix
the encoding is trivial. We simply map $\ket{\psi} $ to $\ket{0}^{\otimes^{n-k}} \ket{\psi}$.
The associated encoded $\ol{X}$ and $\ol{Z}$ operators are given by $(0I_k| 0 0 ) $ and 
$(0 0 |0 I_k)$ respectively. Here we give a sketch of the method for the binary case,
the reader can refer to \cite{grassl03} for details. 
Assume that the stabilizer matrix is given by $S$. 
Then we shall transform it into $(00|I_{n-k}0)$ using the following 
sequence of operations. 
\begin{eqnarray}
(X|Z)  \mapsto (I_{n-k}0|0) \mapsto (00|I_{n-k}0).	
	\label{eq:conjSteps}
\end{eqnarray}
This can be accomplished through the action of $H=\left[\begin{smallmatrix}1&1\\1&-1\end{smallmatrix}\right]$, 
$P=\left[\begin{smallmatrix}1&0\\0&i\end{smallmatrix}\right]$  
and CNOT gates on the Pauli group under conjugation.
The $H$ gate acting on the $i$th qubit on 
$(a_1,\ldots, a_n|b_1,\ldots,b_n)$
transforms it as
\begin{eqnarray}
(a_1,\ldots, a_n|b_1,\ldots,b_n)\stackrel{H_i}{\mapsto}
(a_1,\ldots,\mathbf{b_i},\ldots, a_n|b_1,\ldots,\mathbf{ a_i}, \ldots, b_n).
\end{eqnarray}
These modified entries have been highlighted for convenience.
The phase gate $P$ on the  $i$th qubit transforms $(a_1,\ldots, a_n|b_1,\ldots,b_n)$
as
\begin{eqnarray}
(a_1,\ldots, a_n|b_1,\ldots,b_n)\stackrel{P_i}{\mapsto}
(a_1,\ldots,\mathbf{a_i},\ldots, a_n|b_1,\ldots,\mathbf{ a_i+b_i}, \ldots, b_n).
\end{eqnarray}
We denote the CNOT gate  with the control on the
$i$th qubit and the target on the $j$th qubit by $\text{CNOT}^{i,j}$. The action of the 
$\textup{CNOT}^{i,j}$ gate on $(a_1,\ldots, a_n|b_1,\ldots,b_n)$ is to transform it to 
\begin{eqnarray}
(a_1,\ldots,a_{j-1},\mathbf{ a_j+a_i} ,a_{j+1}\ldots,a_n|b_1,\ldots,b_{i-1},\mathbf{b_{i}+b_{j}},b_{i+1},\ldots, a_n).
\end{eqnarray}
Note that the $j$th entry is changed in the $X$ part while the $i$th entry is changed in the
$Z$ part. For example, consider 
\begin{eqnarray*}
(1,0,0,1,0|0,1,1,0,0) \stackrel{\text{CNOT}^{1,4}}{\mapsto} (1,0,0,\mbf{0},0|0,1,1,0,0),\\
(1,0,0,1,0|0,1,1,1,0) \stackrel{\text{CNOT}^{1,4}}{\mapsto} 
(1,0,0,\mbf{0},0|\mbf{1},1,1,1,0).
\end{eqnarray*}
Based on the action of these three gates we have the following lemmas to transform
error operators.
\begin{lemma}\label{lm:qubitConj}
Assume that we have a error operator of the form $(a_1,\ldots, a_n|b_1,\ldots, b_n)$. Then
we apply the following gates on the $i$th qubit to transform the stabilizer, transforming
$(a_i,b_i)$ to $(\alpha,\beta)$ as per the following table. 
\medskip
\begin{center}
\begin{tabular}{c|c|c}
	$(a_i,b_i)$ & Gate& $(\alpha,\beta)$ \\ \hline
	(0,0) & $I$ & (0,0)\\
	(0,1) & $H$ & (1,0)\\
	(1,0) & $I$ & (1,0)\\
	(1,1) & $P$ & (1,0)
\end{tabular}
\end{center}
Let $\bar{x}$ denote $1+x \bmod 2$, then the transformation to $(a_1,\ldots, a_n|0,\ldots,0)$ is achieved by 
$$
\bigotimes_{i=1}^n H^{\bar{a}_ib_i} P^{a_ib_i}.
$$
\end{lemma}
For example, consider the following generator $(1,0,0,1,0|0,1,1,1,0)$. This can be transformed to
$(1,1,1,1,0|0,0,0,0,0)$ by the application of $I\otimes H \otimes H \otimes P \otimes I $. 

\begin{lemma}\label{lm:cnotConj}
Let $e$ be an error operator of the form $(a_1,\ldots,a_i=1,\ldots, a_n|0,\ldots, 0)$. 
Then $e$ can be transformed to $(0,\ldots,0, a_i=1,0,\ldots, 0 |0,\ldots, 0)$ by
$$\prod_{j=1, i\neq j}^n \left[\textup{CNOT}^{i,j}\right]^{a_j}.$$
\end{lemma}
As an example consider $(1,1,1,1,0|0,0,0,0,0)$, this can be transformed to 
$(0,1,0,0,0|0,0,0,0,0)$ by 
$$
\textup{CNOT}^{2,1} \cdot  \textup{CNOT}^{2,3} \cdot \textup{CNOT}^{2,4}.
$$

The first step involves making the $Z$ portion of the stabilizer matrix all zeros. This
is achieved by single qubit operations consisting of $H$ and $P$
performed on each row one by one.

Note that we must also modify the other rows of the stabilizer matrix according to the action of the
gates applied. 

Once we have a row of stabilizer matrix in the form $(a|0)$, where $a$ is nonozero we can transform
it to the form $(0,\ldots,0,a_i=1,0,\ldots,0|0)$ by using CNOT gates. 
Thus it is easy to transform $(X|Z)$ to $(I_{n-k}0|0)$ using CNOT, $P$ and $H$ gates. 
The final transformation to $(0|I_{n-k}0)$ is achieved by using $H$ gates on the first $n-k$ qubits.
At this point the stabilizer matrix has been transformed to a trivial stabilizer matrix which 
stabilizes the state $\ket{0}^{\otimes^{n-k}}\ket{\psi}$. The encoded operators 
are $(0I_k|0)$ and $(0|0I_k)$. Let $T$ be the sequence of gates applied to transform the 
stabilizer matrix to the trivial stabilizer matrix. Then $T$  applied in the 
reverse order to $\ket{0}^{\otimes^{n-k}}\ket{\psi}$ gives the encoding circuit 
for the stabilizer code. 

\bigskip

Now we shall use the conjugation method to encode the subsystem codes. The main difference is 
that instead of considering just the stabilizer we need to consider the entire gauge group.
Let the gauge group be 
$G= \langle S, G_Z, G_X \rangle$,
where $G_Z=\langle  z_{s+1},\ldots,z_{s+r}\rangle$, and 
$G_X=\langle  x_{s+1},\ldots,x_{s+r}\rangle$.
The idea is to transform the gauge group as follows. 
\begin{eqnarray}
G= \left[\begin{array}{c}S \\ \hline G_Z \\ \hline G_X \end{array}\right]
\mapsto \left[ \begin{array}{ccc|ccc}
0&0&0&I_s&0&0\\ \hline
0&0&0&0&I_r&0\\ \hline
0&I_r&0&0&0&0 \end{array} \right].\label{eq:GnormForm}
\end{eqnarray}
At this point the gauge group has been transformed to a group with trivial stabilizer and
trivial encoded operators for the gauge qubits and the encoded qubits. The sequence of
gates required to achieve this transformation in the reverse order will encode the state
$\ket{0}^{\otimes^{s}}\ket{\phi}\ket{\psi}$. The state $\ket{\phi}$ corresponds to the
gauge qubits and it can be initialized to any state, while $\ket{\psi}$ corresponds to the
input. 

\begin{algorithm}[H]
\caption{{\ensuremath{\mbox{\scshape Encoding subsystem codes -- conjugation method}}}}\label{alg:subsysEncConj}
\begin{algorithmic}[1]
\REQUIRE Gauge group, $G= \langle S, G_Z, G_X \rangle$,
where $G_Z=\langle  z_{s+1},\ldots,z_{s+r}\rangle$, and 
$G_X=\langle  x_{s+1},\ldots,x_{s+r}\rangle$
 and stabilizer, $S =\langle z_1,\ldots, z_{n-k-r}\rangle$ of the $[[n,k,r,d]]$ subsystem code.

\ENSURE $[x_i,x_j] =[z_i,z_j] =0$; $[x_i,z_j ]=2x_iz_i \delta_{ij}$
\medskip
\STATE Assume that $G$ is the following form
$$G= \left[\begin{array}{c}S \\ \hline G_Z \\ \hline G_X \end{array}\right]
$$
\FORALL{$i=1$ to $s+r$}
\STATE  Transform $z_i$ to $z_i'=(a_1,\ldots,a_n|0,\ldots,0)$ using Lemma~\ref{lm:qubitConj}
\STATE Transform $z_i'$ to $(0,\ldots, a_i=1,\ldots,0|0)$ using Lemma~\ref{lm:cnotConj}
\STATE Perform Gaussian elimination on column $i$ for rows $j>i$
\ENDFOR
\STATE Apply $H$ gate on each qubit  $i=1$ to $i=s+r$
\FORALL{$i=s+1$ to $s+r$}
\STATE  Transform $x_i$ to $x_i'=(a_1,\ldots,a_n|0,\ldots,0)$ using Lemma~\ref{lm:qubitConj}
\STATE Transform $x_i'$ to $(0,\ldots, a_i=1,\ldots,0|0)$ using Lemma~\ref{lm:cnotConj}
\STATE Perform Gaussian elimination on column $i$ for rows $j>i$
\ENDFOR
\end{algorithmic}
\end{algorithm}

In the above algorithm, we assume that whenever a row is transformed according to Lemma~\ref{lm:qubitConj}
or \ref{lm:cnotConj}, all the other rows are also transformed according to the transformation applied.

\textbf{Correctness of Algorithm~\ref{alg:subsysEncConj}.} 
The correctness of the algorithm is 
straightforward. As $G$ has full rank of $n-k+r$, for each row of $G$, we will be able to find some
nonzero pair $(a,b)$ so that the the transformation in lines 2--6 can be achieved. When $S$ and $G_Z$
are in the form $(0|I_{s+r}0)$, the rows in $G_X$ are in the form 
$$
\left[\begin{array}{ccc|ccc}0&A&B&0&0&D\end{array} \right]. 
$$
The zero columns of $G_X$ are consequence of the requirement to satsify the commutation
relations with (transformed) $S$ and $G_Z$. For instance, 
The first $n-k-r$ are all zero because
they must commute with $(0|I_s 0)$, the elements of the transformed stabilizer.
The submatrix $A$ must have rank $r$, otherwise at this point one of the rows of $G_X$ commutes with
all the rows of $G_Z$ and the condition that we have there are $r$ hyperbolic pairs 
is violated. It is possible therefore to transform $A$  to the form $(0I_r0|0)$. It cannot be any other
form because then we would not have the $r$ hyperbolic pairs. 
The applied transformations transform $G$ to the form given in equation~(\ref{eq:GnormForm}).
The encoded operators for  this gauge group are clearly $(0I_k|0)$ and $(0|0I_k)$. 
We conclude with a simple example that illustrates the process. 
\begin{example}
To compare with the standard form method, we consider
the $[[4,1,1,2]]$ code again. Let the gauge group $G$,
stabilizer $S$ and encoded operators given by $L$.
\begin{eqnarray*}
S&=&\left[ \begin{array}{cccc}X&X&X&X\\Z&Z&Z&Z 
\end{array}\right] =\left[ \begin{array}{c}z_1\\z_2\end{array}\right],\\
G&=&\left[ \begin{array}{cccc} 
X&X&X&X\\Z&Z&Z&Z \\  \hline
I&I&Z&Z\\I&X&I&X 
\end{array}\right] = \left[ \begin{array}{c}  z_1\\z_2\\\hline x_3\\z_3\end{array}\right].
\end{eqnarray*} 
In matrix form $G$ can be written as 
\begin{eqnarray*}
G&=&\left[ \begin{array}{cccc|cccc} 
1&1&1&1&0&0&0&0\\0&0&0&0&1&1&1&1 \\  \hline
0&0&0&0&0&0&1&1\\
0&1&0&1&0&0&0&0
\end{array}\right].
\end{eqnarray*} 
The transformations consisting of $T_1=\textup{CNOT}^{1,2}\textup{CNOT}^{1,3}\textup{CNOT}^{1,4}$ followed by 
$T_2=I\otimes H\otimes H\otimes H$ maps $G$ to 
\begin{eqnarray*}
\stackrel{T_1}{\mapsto}\left[ \begin{array}{cccc|cccc} 
1&0&0&0&0&0&0&0\\0&0&0&0&0&1&1&1 \\  \hline
0&0&0&0&0&0&1&1\\
0&1&0&1&0&0&0&0
\end{array}\right] \stackrel{T_2}{\mapsto}
\left[ \begin{array}{cccc|cccc} 
1&0&0&0&0&0&0&0\\0&1&1&1&0&0&0&0 \\  \hline
0&0&1&1&0&0&0&0\\
0&0&0&0&0&1&0&1
\end{array}\right].
\end{eqnarray*} 
Now transform the second row using $T_3=\textup{CNOT}^{2,3}\textup{CNOT}^{2,4}$. Then 
transform using $T_4=\textup{CNOT}^{4,3}$. We get
\begin{eqnarray*}
\stackrel{T_3}{\mapsto}\left[ \begin{array}{cccc|cccc} 
1&0&0&0&0&0&0&0\\0&1&0&0&0&0&0&0 \\  \hline
0&0&1&1&0&0&0&0\\
0&0&0&0&0&0&0&1
\end{array}\right] \stackrel{T_4}{\mapsto}
\left[ \begin{array}{cccc|cccc} 
1&0&0&0&0&0&0&0\\0&1&0&0&0&0&0&0 \\  \hline
0&0&0&1&0&0&0&0\\
0&0&0&0&0&0&0&1
\end{array}\right].
\end{eqnarray*} 
Applying $T_5=H\otimes H \otimes I \otimes H $ gives us 
\begin{eqnarray*}
\stackrel{T_5}{\mapsto}\left[ \begin{array}{cccc|cccc} 
0&0&0&0&1&0&0&0\\0&0&0&0&0&1&0&0 \\  \hline
0&0&0&1&0&0&0&0\\
0&0&0&0&0&0&0&1
\end{array}\right].
\end{eqnarray*} 
We could have chosen $T_5= H\otimes H\otimes I \otimes I $, since the effect of 
$H$ on the fourth qubit is trivial. The complete circuit is given as 
\begin{figure}[htb]
\[
\Qcircuit @C=1.4em @R=1.2em {
\lstick{\ket{0}}&\gate{H}&\qw &\qw&\qw&\ctrl{3}&\qw\\
\lstick{\ket{0}}&\gate{H}&\qw&\ctrl{2}&\gate{H}&\targ&\qw\\
\lstick{\ket{\psi}}&\qw&\targ&\targ&\gate{H}&\targ&\qw\\
\lstick{\ket{g}}&\gate{H}&\ctrl{-1}&\targ&\gate{H}&\targ&\qw
}
\]
\caption{Encoding $[[4,1,1,2]]$ code by conjugation method}
\label{fig:conjEnc4112a}
\end{figure}

By switching the target and control qubits of the CNOT gates in $T_3$ and $T_4$
we can show that this circuit is equivalent to circuit shown in 
Figure~\ref{fig:conjEnc4112b}.
\begin{figure}[htb]
\[
\Qcircuit @C=1.4em @R=1.2em {
\lstick{\ket{0}}&\gate{H}&\qw &\qw&\qw&\ctrl{3}&\qw\\
\lstick{\ket{0}}&\qw&\qw&\targ&\targ&\targ&\qw\\
\lstick{\ket{\psi}}&\gate{H}&\ctrl{1}&\ctrl{-1}&\qw&\targ&\qw\\
\lstick{\ket{g}}&\qw&\targ&\qw&\ctrl{-2}&\targ&\qw \gategroup{3}{2}{4}{3}{.7em}{--}
}
\]
\caption{Encoding $[[4,1,1,2]]$ code by conjugation method}\label{fig:conjEnc4112b}
\end{figure}

It is instructive to compare the circuit in Figure~\ref{fig:conjEnc4112b} with 
the one given earlier in Figure~\ref{fig:stdFormEncEx1}. The dotted lines 
show the additional circuitry.
Since the gauge qubit can be initialized to any state, we can initialize
$\ket{g}$ to $\ket{0}$, which then gives the following logical states for the code. 
\begin{eqnarray}
\ket{\ol{0}}&=&\ket{0000}+\ket{1111}+\ket{0011}+\ket{1100},\\
\ket{\ol{1}}&=&\ket{0000}+\ket{1111}-\ket{0011}-\ket{1100}.
\end{eqnarray}
It will be observed that $IIXX$ acts as the logical $Z$ operator while $IZIZ$ acts as the
logical $X$ operator. We could flip these logical operators by absorbing the $H$ gate into
$\ket{\psi}$. If we additionally initialize $\ket{g}$ to $\ket{0}$, we will see that
the two CNOT gates on the second qubit can be removed. The circuit then simplifies to 
the circuit shown in Figure~\ref{fig:conjEnc4112c}.

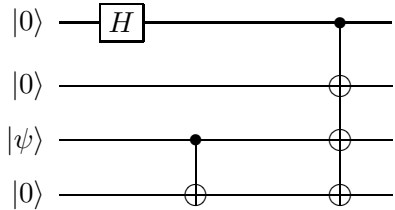
\begin{figure}[htb]
\[
\Qcircuit @C=1.4em @R=1.2em {
\lstick{\ket{0}}&\gate{H}&\qw &\qw&\qw&\ctrl{3}&\qw\\
\lstick{\ket{0}}&\qw&\qw&\qw&\qw&\targ&\qw\\
\lstick{\ket{\psi}}&\qw&\ctrl{1}&\qw&\qw&\targ&\qw\\
\lstick{\ket{0}}&\qw&\targ&\qw&\qw&\targ&\qw 
}
\]
\caption{Encoding $[[4,1,1,2]]$ code by conjugation method -- optimized}
\label{fig:conjEnc4112c}
\end{figure}

This is precisely, the same circuit that we had arrived earlier in Figure~\ref{fig:stdFormEncEx1Opt}
using the standard
form method. 
\end{example}
The preceding example provides additional evidence in the direction that 
it is better to initialize the gauge qubits to zero
and avoid the encoding operators on them.

\paragraph{Conclusions.}
In this paper, we have demonstrated that the subsystem codes can be encoded 
using the techniques used for stabilizer codes. In particular, we have
considered two methods for encoding stabilizer codes -- the standard form
method and the conjugation method. While the standard form method explored
here required us to initialize the gauge qubits to zero, 
it admits two two variants and seems to have
the potential for lower complexity; the exact gains being determined by
the actual codes under consideration. The conjugation method allows us
to initialize the gauge qubits to any state. The disadvantage seems to be
the increased complexity of encoding. It must be emphasized that the standard
form method is equivalent to the conjugation method and it is certainly
possible to use this method to encode subsystem codes so that the gauge
qubits can be initialized to arbitrary states. However, it appears to 
be a little more cumbersome and for this reason we have not investigated
this in this paper. There is yet another method for encoding stabilizer codes
based on the teleportation due to Knill. We expect that gauge qubits can 
be exploited even in this method to reduce its complexity. It would be
interesting to investigate fault tolerant encoding schemes for subsystem codes
exploiting the gauge qubits.

\section*{Appendix}
\paragraph{The logical states of a stabilizer code.}
We assume that our basis input states are of the 
form $\ket{0}^{\otimes^{n-k}}\ket{\alpha_1\ldots\alpha_k}$, where $\alpha_i\in \{ 0, 1 \}$. 
Clearly, we have freedom in the choice of the states into which each of these
states are encoded to. Additionally, we have freedom in the choice of
the encoded operators though they are not entirely unrelated. 
Perhaps, this is best illustrated through an example.
Let us consider Shor's $[[9,1,3]]_2$ code. 
A choice of the logical states for this code is
\begin{eqnarray*}
\ket{\ol{0}}&=&(\ket{000}+\ket{111})(\ket{000}+\ket{111})(\ket{000}+\ket{111}),\\
\ket{\ol{1}}&=&(\ket{000}-\ket{111})(\ket{000}-\ket{111})(\ket{000}-\ket{111}).
\end{eqnarray*}
For this choice of the encoded states the logical $Z$ operator is 
$X^{\otimes^9}$ and the logical $X$ operator is $Z^{\otimes^9}$. 
On the other hand, let us see what happens if we choose the logical states as follows:
\begin{eqnarray*}
\ket{\ol{0}}&=&\ket{000000000}+\ket{000111111}+\ket{111000111}+\ket{111111000},\\
\ket{\ol{1}}&=&\ket{111111111}+\ket{111000000}+\ket{000111000}+\ket{000000111}.
\end{eqnarray*}
In this case the encoded $X$ operator is $X^{\otimes^9}$ and encoded $Z$ operator is
$Z^{\otimes^9}$; they are flipped with respect to the previous choice!

So it becomes apparent that the assignment of the encoded operators as logical 
$Z$ or $X$ is flexible and it seems to depend on the choice of the logical states.
But are we free to choose any basis of the codespace as the encoded 
logical states. We can show that this cannot be. 
For instance let us choose the logical zero state to be a superposition of the 
previous two assignments. Then we have 
\begin{eqnarray*}
\ket{\ol{0}}&=&(\ket{000}+\ket{111})(\ket{000}+\ket{111})(\ket{000}+\ket{111})\\
&+&\ket{000000000}+\ket{000111111}+\ket{111000111}\\&+&\ket{111111000}.
\end{eqnarray*}

The possibilities for the logical $Z$ operator\footnote{Including scalar multiples of $i$
will not change our conclusions.} are $\pm X^{\otimes^9}$, $\pm Z^{\otimes^9}$,
$\pm X^{\otimes^9} Z^{\otimes^9}$. But for none of these operators we have 
$\ol{Z}\ket{\ol{0}}=\ket{\ol{0}}$. As these are the only possible encoded operators (modulo the stabilizer which acts trivially in any case), this is not a valid choice for $\ket{\ol{0}}$.
This raises the question what are all the possible valid choices for the logical states. 
Let us look at yet another choice of logical states. 
\begin{eqnarray*}
\ket{\ol{0}}&=&(\ket{000}-\ket{111})(\ket{000}-\ket{111})(\ket{000}-\ket{111}),\\
\ket{\ol{1}}&=&(\ket{000}+\ket{111})(\ket{000}+\ket{111})(\ket{000}+\ket{111}).
\end{eqnarray*}
In this case, the encoded $Z$ and $X$ operators 
are $-X^{\otimes^9}$ and $Z^{\otimes^9}$ respectively. 
This gives us a clue as to the possible logical all zero states
for a given stabilizer code. The all zero logical state is the state in the 
code space that is fixed by the stabilizer and the logical $Z$ operators. 
Assuming that $S$ is the stabilizer and $C_{\mc{P}_n}(S)$, its centralizer, we can 
can pick any $k$ independent commuting generators in $C_{\mc{P}_n}(S)\setminus S Z({\mc{P}_n})$ as $Z$ operators. 
Hence, we have the following lemma.

\begin{lemma}\label{lm:logicalZero}
Let $S$ be the stabilizer of an $[[n,k,d]]_2$ stabilizer code.
If $L \le C_{\mc{P}_n}(S)$ is any subgroup generated by 
$n$ commuting generators such that $L\cap Z({\mc{P}_n})=I$ and $S\le L$, then 
the state stabilized by $L$ is a valid logical all zero state for the stabilizer 
code defined by $S$.
\end{lemma}

The implicit choice of $\ket{\ol{0}}$ made in Lemma~\ref{lm:stabStdForm} 
(by picking the encoded $Z$ operators, at least the representatives) is convenient in
the sense it allows us to speak of a canonical $\ket{\ol{0}}$
without ambiguity. This $\ket{\ol{0}}$ can be conveniently identified
with the state $P\ket{0}^{\otimes^n}$, where it will be recalled that $P$
is the projector for the stabilizer code given as
\begin{eqnarray}
P &=& \frac{1}{|S|}\sum_{M\in S}  M.
\end{eqnarray}

\def\cprime{$'$}

\end{document}